\documentstyle[preprint,aps]{revtex}
\input epsf              % uncomment to use epsf
\begin{document}
\begin{flushright}
{\bf Fermilab-Pub-97/363-E}
\end{flushright}
\vskip 0.5cm
\begin{center} 
\Large \bf $Z\gamma$ Production in $p\bar{p}$ Collisions at 
$\sqrt{s}=1.8$ TeV and  Limits on Anomalous $ZZ\gamma$ and $Z\gamma\gamma$ 
Couplings
\end{center}
\vskip 1.0cm
\centerline{The D\O \ Collaboration$^*$}
\centerline{\it Fermi National Accelerator Laboratory, Batavia, IL 60510}
\centerline{(October 29, 1997)}
\vskip 1.5cm
\centerline{\Large Abstract}
\vskip 1.0cm

 We present a study of $Z \gamma + X$ production in $p\bar{p}$ collisions at 
 $\sqrt{s}=1.8$ TeV from 97 (87) pb$^{-1}$ of data collected
 in the $ee\gamma$ $(\mu \mu \gamma)$ decay channel with the 
 D{\O} detector at Fermilab. The event yield and kinematic characteristics are
 consistent with the Standard Model predictions.  
%We obtain 95\% CL limits 
% on anomalous  $ZZ\gamma$ and $Z\gamma\gamma$ couplings, 
% $|h^{Z}_{30}| < 1.31$, $|h^{Z}_{40}| < 0.26$, $|h^{\gamma}_{30}| < 1.36$,
% and $|h^{\gamma}_{40}| < 0.26$ for a form factor scale $\Lambda=500$ GeV.
 We obtain limits on anomalous  $ZZ\gamma$ and $Z\gamma\gamma$ couplings
    for form factor scales $\Lambda = 500$ GeV and $\Lambda = 750$ GeV.
% Combining these with our previous results yields 
 Combining this analysis with our previous results yields 95\% CL limits
 $|h^{Z}_{30}| < 0.36$, $|h^{Z}_{40}| < 0.05$, $|h^{\gamma}_{30}| < 0.37$, 
 and $|h^{\gamma}_{40}| < 0.05$ for a form factor scale 
 $\Lambda=750$ GeV.

\begin{flushleft}
------ \\
$^*$ Authors listed on the following page. \\
Submitted to Physical Review Letters.
\end{flushleft}
\global\advance\count0 -1
\clearpage

%\input pslogos
%\draft         % print PACS numbers
\title{$Z\gamma$ Production in $p\bar{p}$ Collisions at $\sqrt{s}=1.8$ TeV
and Limits on Anomalous $ZZ\gamma$ and $Z\gamma\gamma$ Couplings}

% LIST_OF_AUTHORS.TEX                 09/18/97           
%
\author{                                                                      
%% names begin here                                                           
B.~Abbott,$^{30}$                                                             
M.~Abolins,$^{27}$                                                            
B.S.~Acharya,$^{45}$                                                          
I.~Adam,$^{12}$                                                               
D.L.~Adams,$^{39}$                                                            
M.~Adams,$^{17}$                                                              
S.~Ahn,$^{14}$                                                                
H.~Aihara,$^{23}$                                                             
G.A.~Alves,$^{10}$                                                            
E.~Amidi,$^{31}$
% ...
\\                                                                            
\vskip 0.50cm                                                                 
\centerline{(D\O\ Collaboration)}                                             
\vskip 0.50cm                                                                 
}                                                                             
\address{                                                                     
\centerline{$^{1}$Universidad de los Andes, Bogot\'{a}, Colombia}             
\centerline{$^{2}$University of Arizona, Tucson, Arizona 85721}               
\centerline{$^{3}$Boston University, Boston, Massachusetts 02215}             
\centerline{$^{4}$Brookhaven National Laboratory, Upton, New York 11973}      
\centerline{$^{5}$Brown University, Providence, Rhode Island 02912}           
\centerline{$^{6}$Universidad de Buenos Aires, Buenos Aires, Argentina}       
\centerline{$^{7}$University of California, Davis, California 95616}          
\centerline{$^{8}$University of California, Irvine, California 92697}         
\centerline{$^{9}$University of California, Riverside, California 92521}      
\centerline{$^{10}$LAFEX, Centro Brasileiro de Pesquisas F{\'\i}sicas,        
                  Rio de Janeiro, Brazil}                                     
\centerline{$^{11}$CINVESTAV, Mexico City, Mexico}                            
\centerline{$^{12}$Columbia University, New York, New York 10027}             
\centerline{$^{13}$Delhi University, Delhi, India 110007}                     
\centerline{$^{14}$Fermi National Accelerator Laboratory, Batavia,            
                   Illinois 60510}                                            
\centerline{$^{15}$Florida State University, Tallahassee, Florida 32306}      
\centerline{$^{16}$University of Hawaii, Honolulu, Hawaii 96822}              
\centerline{$^{17}$University of Illinois at Chicago, Chicago,                
                   Illinois 60607}                                            
\centerline{$^{18}$Indiana University, Bloomington, Indiana 47405}            
\centerline{$^{19}$Iowa State University, Ames, Iowa 50011}                   
\centerline{$^{20}$Korea University, Seoul, Korea}                            
\centerline{$^{21}$Kyungsung University, Pusan, Korea}                        
\centerline{$^{22}$Institute of Nuclear Physics, Krak\'ow, Poland}            
\centerline{$^{23}$Lawrence Berkeley National Laboratory and University of    
                   California, Berkeley, California 94720}                    
\centerline{$^{24}$Louisiana Tech University, Ruston, Louisiana 71272}        
\centerline{$^{25}$University of Maryland, College Park, Maryland 20742}      
\centerline{$^{26}$University of Michigan, Ann Arbor, Michigan 48109}         
\centerline{$^{27}$Michigan State University, East Lansing, Michigan 48824}   
\centerline{$^{28}$Moscow State University, Moscow, Russia}                   
\centerline{$^{29}$University of Nebraska, Lincoln, Nebraska 68588}           
\centerline{$^{30}$New York University, New York, New York 10003}             
\centerline{$^{31}$Northeastern University, Boston, Massachusetts 02115}      
\centerline{$^{32}$Northern Illinois University, DeKalb, Illinois 60115}      
\centerline{$^{33}$Northwestern University, Evanston, Illinois 60208}         
\centerline{$^{34}$University of Notre Dame, Notre Dame, Indiana 46556}       
\centerline{$^{35}$University of Oklahoma, Norman, Oklahoma 73019}            
\centerline{$^{36}$University of Panjab, Chandigarh 16-00-14, India}          
\centerline{$^{37}$Institute for High Energy Physics, 142-284 Protvino,       
                   Russia}                                                    
\centerline{$^{38}$Purdue University, West Lafayette, Indiana 47907}          
\centerline{$^{39}$Rice University, Houston, Texas 77005}                     
\centerline{$^{40}$Universidade do Estado do Rio de Janeiro, Brazil}          
\centerline{$^{41}$University of Rochester, Rochester, New York 14627}        
\centerline{$^{42}$CEA, DAPNIA/Service de Physique des Particules,            
                   CE-SACLAY, Gif-sur-Yvette, France}                         
\centerline{$^{43}$Seoul National University, Seoul, Korea}                   
\centerline{$^{44}$State University of New York, Stony Brook,                 
                   New York 11794}                                            
\centerline{$^{45}$Tata Institute of Fundamental Research,                    
                   Colaba, Mumbai 400005, India}                              
\centerline{$^{46}$University of Texas, Arlington, Texas 76019}               
\centerline{$^{47}$Texas A\&M University, College Station, Texas 77843}       
}                                                                             
%end                                                                          

\maketitle

\begin{abstract}
 We present a study of $Z \gamma + X$ production in $p\bar{p}$ collisions at 
 $\sqrt{s}=1.8$ TeV from 97 (87) pb$^{-1}$ of data collected
 in the $ee\gamma$ $(\mu \mu \gamma)$ decay channel with the 
 D{\O} detector at Fermilab. The event yield and kinematic characteristics are
 consistent with the Standard Model predictions.  
%We obtain 95\% CL limits 
% on anomalous  $ZZ\gamma$ and $Z\gamma\gamma$ couplings, 
% $|h^{Z}_{30}| < 1.31$, $|h^{Z}_{40}| < 0.26$, $|h^{\gamma}_{30}| < 1.36$,
% and $|h^{\gamma}_{40}| < 0.26$ for a form factor scale $\Lambda=500$ GeV.
 We obtain limits on anomalous  $ZZ\gamma$ and $Z\gamma\gamma$ couplings
    for form factor scales $\Lambda = 500$ GeV and $\Lambda = 750$ GeV.
% Combining these with our previous results yields 
 Combining this analysis with our previous results yields 95\% CL limits
 $|h^{Z}_{30}| < 0.36$, $|h^{Z}_{40}| < 0.05$, $|h^{\gamma}_{30}| < 0.37$, 
 and $|h^{\gamma}_{40}| < 0.05$ for a form factor scale 
 $\Lambda=750$ GeV. 
\end{abstract}

%\narrowtext
%\twocolumn

%\section{Introduction}

Studies of vector boson pair production and measurements of the trilinear 
gauge boson couplings provide important tests of the Standard Model (SM) 
of electroweak interactions.  
The SM predicts no tree-level couplings between the $Z$ boson and the 
photon. Observation of such couplings would indicate the 
presence of new physical phenomena.  
Recent limits on the $ZZ\gamma$ and 
$Z\gamma\gamma$ coupling parameters have been obtained by CDF~\cite{CDF95},
L3~\cite{L395}, DELPHI~\cite{D380} and D\O~\cite{D095,D0vvg}.

In the SM, the $\ell^{+}\ell^{-}\gamma$ final state can be 
produced via radiative decays of the $Z$ boson or by production of a 
boson pair via $t$- or $u$-channel quark exchange. 
The former process is the dominant source of events with small 
opening angle between
the photon and charged lepton and for events with a low value of photon 
transverse energy, $E_T^{\gamma}$.  Events produced by the 
latter process have lepton-pair invariant mass, $m_{\ell\ell}$, close to $M_Z$ 
and three-body invariant mass, $m_{\ell\ell\gamma}$, larger than $M_Z$. 
Anomalous $ZZ\gamma$ or $Z\gamma\gamma$ couplings would enhance the cross 
section for $Z\gamma$ production, particularly for high-$E_T$ photons, 
relative to the SM expectations. 

A study of $Z\gamma$ production and a search for anomalous $Z\gamma$ couplings 
has been performed using the reactions $p\bar{p} \rightarrow ee\gamma X$ and 
$\mu\mu\gamma X$ at $\sqrt{s} = 1.8$ TeV in data collected with the D{\O} 
detector at Fermilab during the 1993-1995 Tevatron run.  
These data correspond to an 
integrated luminosity of $97 \pm 5$ $(87\pm 5)$ pb$^{-1}$ in the 
$ee\gamma$ $(\mu\mu\gamma)$ channels.
This study is 
complementary to that of Ref.~\cite{D0vvg}, which sets limits on 
anomalous $ZV\gamma$ ($V = Z, \gamma$) couplings using a fit to the 
$E_T^{\gamma}$ spectrum
in events analyzed with the $Z\rightarrow \nu\bar{\nu}$ hypothesis.  
The sensitivities to anomalous couplings are equivalent based on 
the expected event yields and $E_T^{\gamma}$ spectra.  
The backgrounds are dissimilar and the 
signal-to-background ratio is much higher in the charged-lepton analysis. 
Also, the kinematic characteristics of the charged-lepton events can be 
studied in detail.

The results of the search for anomalous couplings 
are presented within the formalism of
Ref.~\cite{Baur93}, which assumes only that any possible
trilinear $ZV\gamma$  coupling must obey Lorentz and
gauge invariance.  In this formalism, the most general $ZV\gamma$ 
vertex contains four undetermined coupling parameters 
$h^{V}_{i}$ $(i=1,\ldots ,4)$.  
Terms proportional to $h^{V}_{1}$ and $h^{V}_{2}$ 
in the scattering amplitudes are CP-odd, while those proportional to
$h^{V}_{3}$ and $h^{V}_{4}$ are CP-even.  To ensure partial wave
unitarity at high energies, a form factor ansatz 
$h^{V}_{i}(\hat{s}) = h^{V}_{i0}/(1+\hat{s}/\Lambda^{2})^{n_{i}}$ is
used \cite{Baur93}, where $\sqrt{\hat{s}}$ is the parton
center-of-mass energy, $h^{V}_{i0}$ is the value of $h^{V}_{i}$ in the
low-energy limit $\hat{s} = 0$, $\Lambda$ is a mass scale, and
$n_{i}$ is the form factor power.  
Form factor powers of $n_{1}=n_{3} = 3$ and $n_{2}=n_{4} = 4$ were used.
These choices of $n_i$ 
 provide the terms in the amplitude proportional to $h^{V}_{i}$
with same high energy behavior.

%\section{Apparatus}

The D{\O} detector, described in detail in Ref.~\cite{D0NIM},
consists of three main systems: the inner tracker, the calorimeter, and the 
muon systems. A nonmagnetic central tracking system, composed of central and
forward drift chambers, provides directional information for charged
particles and is used in this analysis to discriminate between
electrons and photons and in muon identification. 
Particle energies are measured by a liquid-argon uranium
sampling calorimeter that is divided into three cryostats.
The central calorimeter (CC) covers pseudorapidity $|\eta| < 1.1$, and the 
end calorimeters (EC) cover $1.1 < |\eta| < 4.4$.  
The EM (hadron) calorimeters are divided into four (four to six) layers to 
measure longitudinal shower development.  
Energy resolutions of approximately $\sigma(E)/E = 15\%/\sqrt{E}\oplus0.4\%$ 
($E$ in GeV) are achieved for electrons 
and photons. The muon system consists of magnetized iron toroids with 
one inner and two outer layers of drift tubes and achieves a momentum 
resolution of $\sigma(1/p)=0.18(p-2)/p^2 \oplus 0.003$ with $p$ in GeV/c.

Data were collected with a multi-level trigger system. 
The $ee\gamma$ candidates were required to contain two EM clusters with
$E_{T} > 20$ GeV.  The trigger efficiency was estimated to be
nearly 100\% for events that satisfied the offline $ee\gamma$ 
selection criteria
given in the next paragraph. The $\mu\mu\gamma$ candidates were required 
to have at least one muon within $|\eta|<1$ and $p_T > 8$ GeV/c and to have 
an EM cluster with $E_{T} > 7$ GeV. 
The trigger efficiency ranged from $60\%$ to $90\%$ depending on 
$E_T^{\gamma}$ and on whether 
the event passed the tight or loose muon selection described below.

%\section{Event Selection}

Events which satisfied the trigger requirements were subjected to 
further selection criteria.  Each 
$ee\gamma$ event was required to have
two electron candidates with $E_{T} > 25$ GeV and a photon
candidate with $E_{T} > 10$ GeV within the fiducial region $|\eta|<1.0$ (CC) 
or $1.5<|\eta|<2.5$ (EC).  Of the two electron candidates, one was required 
to have a matching track, and the other was required to have a track or 
drift chamber hits associated with the electromagnetic shower.  
The photon was required to have no matching track and no drift chamber
hits nearby.  

Two samples of $\mu\mu\gamma$ candidates were identified.  The events 
identified using the tight selection criteria were required to have a photon, 
and two isolated muon tracks in the region $|\eta|<1$.  The events 
identified using the loose selection criteria were required to have: a photon; 
an isolated muon track in the region $|\eta|<1$; and a muon 
identified\cite{MTC1,MTC2} by a pattern of isolated energy deposition in the 
longitudinal segments of the hadron calorimeter in the region $|\eta|<2.4$, 
with an azimuth, $\phi$, within 0.4 radians of the direction of the 
missing transverse energy corrected for the $p_T$ of the tracked muon. 
In the tight selection, 
one muon was required to have $p_T>15$ GeV/c and the other to have $p_T>10$ 
GeV/c. In the loose selection, the muon with a track was required to have 
$p_T>15$ GeV/c. 
In both selections the opening angle between the muons was required to 
be between 40 and 160 degrees.
Also, the photon candidate was required 
to be within the fiducial region $|\eta|<1.1$ (CC) or $1.5<|\eta|<2.5$ (EC), 
to have $E_T>10$ GeV, and not to have a matching central 
detector track. 

%Electron and photon candidates were required to (i) have 90\% of their
%shower energies contained in the EM calorimeter compartments, (ii)
%have a longitudinal and transverse shower profile consistent with that
%expected for electrons, and (iii) were required to be isolated from
%activity in the hadronic calorimeter.  The latter requirement is based
%on the the isolation variable ${\cal I} = (E(0.4) - E_{\rm
%EM}(0.2))/E_{\rm EM}(0.2)$, where $E(0.4)$ is the total energy in the
%cone centered on the shower with radius ${\cal R} = 0.4$, and $E_{\rm
%EM}(0.2)$ is the electromagnetic energy in the cone of radius 0.2.
%The isolation requirement for electrons and photons was ${\cal I} <
%0.10$.

An angular separation of ${\cal R}_{\ell\gamma} \equiv \sqrt{\Delta\eta^{2}
+ \Delta\phi^{2}}>0.7$ was required between the photon and 
the electrons or muons.  This reduces the number of radiative $Z \rightarrow
\ell^{+}\ell^{-}\gamma$ decay events in the final sample while maintaining
sensitivity to $ZV\gamma$ couplings.

%\section{Signal Efficiencies}

The efficiencies for the above selection criteria were estimated using
$Z \rightarrow ee$ and $Z\rightarrow \mu\mu$ candidates in the data. 
For electrons, the detection efficiency was measured to be about 80\% when a 
track match was required.  When only drift chamber hits were
required, the efficiency increased to about 90\%.  
Including the geometrical acceptance, the muon tracking and reconstruction 
efficiency was $41\pm 2\%$ for $|\eta|<1.0$, and $80\pm 2\%$  ($64\pm3 \%$) 
for muons identified by the calorimeter 
with $|\eta|<1.1$ $(1.1<|\eta|<2.4)$. 
The overall acceptance of the loose $\mu\mu\gamma$ selection criteria was 
3.2 times greater than that of the tight $\mu\mu\gamma$ 
selection criteria.
The photon efficiency was found to depend on $E_T$ and $\eta$, and ranged from
35\% for EC photons at $E_{T}^{\gamma} = 10$ GeV to approximately 70\%
for CC photons with $E_{T}^{\gamma} > 25$ GeV.  The efficiency of the
veto against photons with drift chamber hits or tracks in close proximity, 
used in the $ee\gamma$ analysis,
ranged from 80\% in the CC to 60\% in the EC.  The energy dependence 
of the photon detection efficiency, due to the effects of the underlying 
event and noise, was estimated from photons simulated with 
{\sc geant}~\cite{geant} superimposed on minimum bias data collected 
during the run.

A parametric detector simulation \cite{Glen96} along with a leading-order MC
event generator \cite{Baur93},
was used to predict the signal as a function of the couplings
$h^{V}_{i0}$.  A K-factor of 1.34~\cite{Baur93} was used to correct the 
predicted cross section for processes not included in the leading-order
calculation.  Additionally, the $\ell\ell\gamma$ system was given a
transverse momentum according to the theoretical prediction for $Z$ boson
production \cite{Ladi94} to simulate kinematic effects\cite{Ohne95} 
not included in the event generator.  
Parton densities were taken from the MRSD${-}^{\prime}$ set~\cite{MRS93}.  
A total theoretical uncertainty of 6\% is
assigned to the signal prediction.  This uncertainty reflects the variation in
predicted signal for $Q^{2}$  scales in the range 
$\hat{s}/4 < Q^{2} < 4\hat{s}$ using recently fitted parton densities.
%The efficiency for the event selection was 5\% (6\%) for the 
%$ee\gamma$ $(\mu\mu\gamma)$ channel.

With an integrated luminosity of 97 (87) pb$^{-1}$, the expected SM $ee\gamma$ 
$(\mu\mu\gamma)$ signal is $13.2 \pm 1.3$ $(16.3\pm 2.0)$ events. 
The contributions to the systematic uncertainty on this 
prediction, listed in Table 1, total 10\% (12\%).

%\section{Backgrounds}

The major source of background in the electron decay 
channel is $Z+ \rm jets$ production with a jet misidentified as a photon.  
Contributions from multijet and direct photon ($\gamma +$ jets) processes in 
which one or more jets are misidentified as an electron or photon 
are smaller but not negligible. 
Similarly, the major background for the muon decay channel is 
$Z+ \rm jets$ production. The sample selected with the loose 
selection criteria also includes substantial background from $W+\rm jets$ with 
a fake muon and a jet misidentified as a photon. 

The 
probability for a jet to be misidentified as a photon was measured from an 
independent sample of multijet events.  After subtracting the expected number 
of direct photons in the sample, the misidentification probability 
$P_{j\rightarrow\gamma}$ was found to depend slightly on $E_T^{\rm jet}$ and 
was estimated to be $\sim 10^{-3}$.  A systematic uncertainty of 25\% 
assigned to $P_{j\rightarrow\gamma}$ accounts for 
the uncertainty in the direct photon fraction of the multijet sample.  
The electron misidentification probability $P_{j\rightarrow e}$ was
measured in a similar way and was found to be about half of 
$P_{j\rightarrow\gamma}$. 
The backgrounds in the $ee\gamma X$ and $\mu\mu\gamma X$ candidate samples were
estimated by weighting $eejX$, $\mu\mu jX$, and $e \gamma j X$ events 
in the parent sample by the appropriate $P_{j\rightarrow\gamma}$ and 
$P_{j\rightarrow e}$ factors.  
The background from events with jets misidentified as electrons or photons 
is $1.81 \pm 0.54$ events for the $ee\gamma$ channel, 
$0.29\pm0.08$ events for the tight $\mu\mu\gamma$ 
channel, and $1.89\pm0.54$ events for the loose $\mu\mu\gamma$ channel.

Contributions from processes such as $Z\gamma \rightarrow
\tau^{+} \tau^{-} \gamma$ and $WZ \rightarrow \ell\ell e\nu$ were
investigated and found to be negligible for the $ee\gamma$ channel and for the 
$\mu\mu\gamma$ channel selected with the tight criteria.
However, the $\mu\mu\gamma$ sample selected with the loose selection criteria
has backgrounds of $1.11\pm 0.30$ events from $W\gamma \rightarrow \mu\nu
\gamma + X$, $0.28\pm0.08$ events from $Z\rightarrow \tau\tau\rightarrow 
\mu e + X$, and $0.013\pm0.002$ events from $WW$ and $t\bar{t} \rightarrow
\mu e + X$, which arise because of a fake muon. The probability for fake 
muons was measured using the $Z\rightarrow ee$ and $Z\rightarrow\mu\mu$ data.

%\section{Results}

In the data 14 (15) $ee\gamma$ $(\mu \mu \gamma )$ candidate events were 
identified. Four of the $\mu\mu\gamma$ events were from 
the tight selection criteria and 11 were from the loose selection criteria.
The predicted total background is $1.81\pm 0.54$ $(3.57 \pm 0.68)$ events 
in the $ee\gamma$ $(\mu\mu\gamma)$ channel. Thus, the measured signal is 
$12.2 \pm 3.8$ $(11.4\pm3.9)$ events. 
The total is consistent with the predictions of the SM, as are the 
contributions from the individual channels.

\begin{figure}[ht]
\epsfxsize = 9.0cm
\epsffile{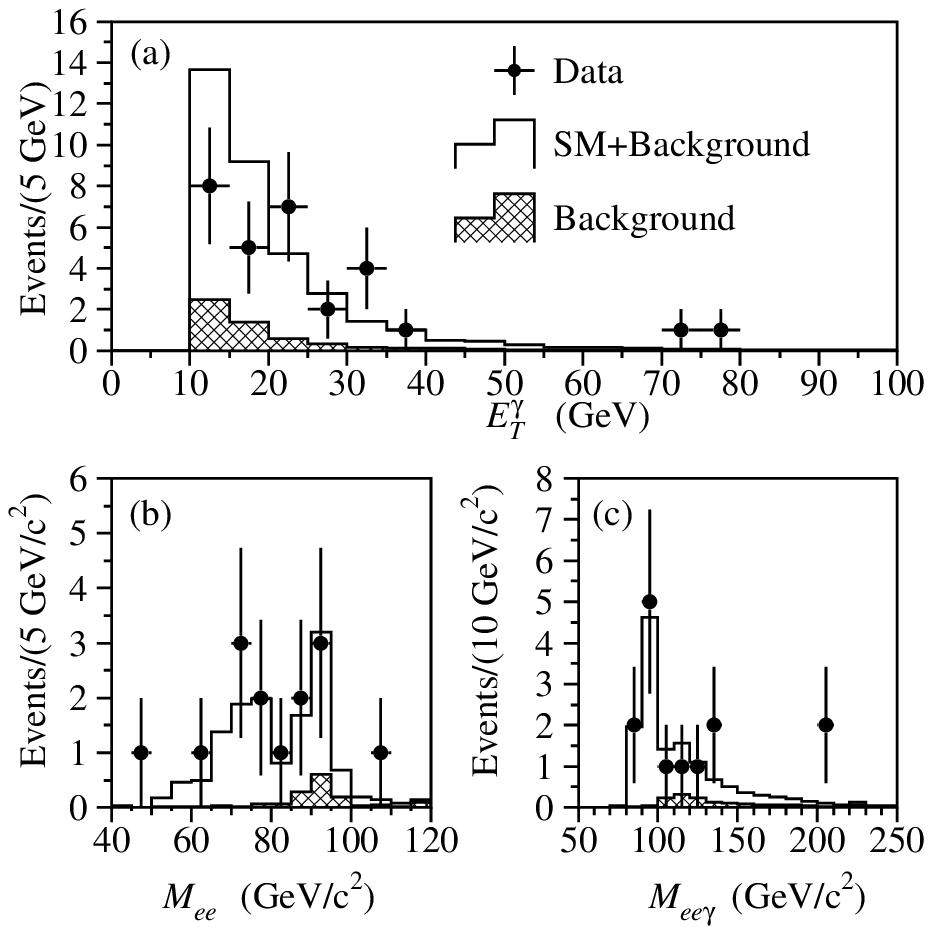}
\caption{Kinematic distributions for candidates and background
estimates:  (a) photon transverse energy for the combined $ee\gamma$ and 
$\mu\mu\gamma$ samples, 
(b) dielectron invariant mass, (c) dielectron-photon invariant mass.}
\label{fig.kin}
\end{figure}

The kinematic distributions of the candidates are shown in
Fig.~\ref{fig.kin}, along with the corresponding background
distributions.  Figure~\ref{fig.kin}(a) shows the $E^{\gamma}_T$ spectrum
of the combined electron and muon channels.
The spectrum is consistent with the expectation of the SM.  
Figures~\ref{fig.kin}(b) and \ref{fig.kin}(c) show the
dielectron invariant mass and dielectron-photon invariant mass, respectively.
Two $ee\gamma$ events were observed with $E^{\gamma}_{T} \approx 75$ GeV, 
dielectron invariant mass $M_{ee} \approx M_{Z}$, and dielectron-photon 
invariant mass $M_{ee\gamma} \approx 200$ GeV/c$^{2}$.  
Assuming SM $Z\gamma$ production, the probability of
observing two or more events with $E^{\gamma}_{T} > 60$ $(70)$ GeV in the 
combined electron and muon channels is 15\% (7.3\%). 
The SM Monte Carlo indicates the 
most likely $ee\gamma$ mass for events with $E^{\gamma}_{T}$ in the range 
70 to 79 GeV is 200 GeV/c$^{2}$. Thus the two events can be understood as a 
fluctuation of SM $Z\gamma$ production.  Note that the dielectron mass 
distribution shows
indications of the predicted two-peaked structure 
induced by the photon $E_T$ threshold and the $e\gamma$ opening angle
selection criteria used to suppress the radiative events. 
The number of $Z\gamma$ production candidates with 
$M_{ee}>83$ GeV/c$^2$ and $M_{ee\gamma}>100$ GeV/c$^2$ is consistent with 
the SM prediction. 
  The plots analogous to 
Figs.~\ref{fig.kin}(b) and \ref{fig.kin}(c) for the muon 
channel show agreement with the SM predictions, but the detailed structure 
seen in the electron channel plots is obscured by the limited
momentum resolution of the muon system. 

\begin{figure}[ht]
\epsfxsize = 10.0cm
\epsffile{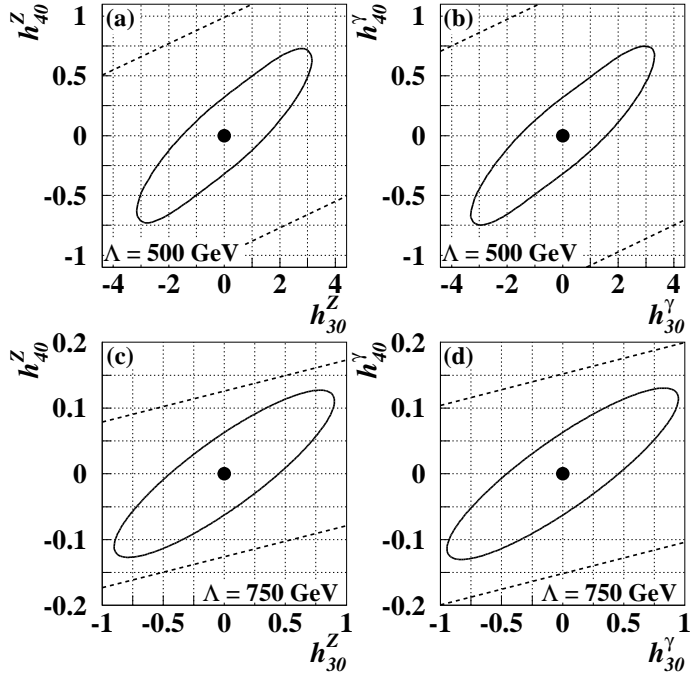}
\caption{Two-dimensional limits (a) on $h^{Z}_{30}$ vs. $h^{Z}_{40}$,  
 and (b) on $h^{\gamma}_{30}$ vs. $h^{\gamma}_{40}$ from the 
 $ee (\mu\mu)\gamma$  analyses and the same, (c) and (d), from combining 
 this analysis with previous
 results from this experiment.  Only the couplings 
 varied in each plot are assumed to be different from the SM values.
 Unitarity limits are indicated by the dashed contours. }
\label{fig.zzg}
\end{figure}

Limits on the $ZV\gamma$ couplings were extracted from the data by
performing an unbinned likelihood fit to the $E^{\gamma}_{T}$
distribution that utilized both the shape of the photon spectrum and
the total event yield.  The likelihood function was convoluted with 
Gaussian probability distributions to account for the systematic
uncertainties.  
With the constraint that only one coupling be nonzero at a time (1D), the 
95\% confidence level (CL) limits are $|h^{Z}_{30}| < 1.31$, 
$|h^{Z}_{40}| < 0.26$, $|h^{\gamma}_{30}| < 1.36$, and 
$|h^{\gamma}_{40}| < 0.26$ for a form factor scale $\Lambda = 500$ GeV. 
Contours for the 95\% CL two-dimensional (2D) limits\cite{2dlim} on the 
CP-even $ZZ\gamma$ and $Z\gamma\gamma$ coupling pairs (where two of the
anomalous couplings are allowed to vary at the same time) are shown in 
Figs.~\ref{fig.zzg}(a) and \ref{fig.zzg}(b). 
With $\Lambda = 750$ GeV, the 1D limits are $|h^{Z}_{30}| < 0.67$,
$|h^{Z}_{40}| < 0.08$, $|h^{\gamma}_{30}| < 0.69$, and 
$|h^{\gamma}_{40}| < 0.08$.  The 2D limits for 
$\Lambda = 750$ GeV are slightly looser than the unitarity constraints.
The limits on the CP-odd couplings are nearly identical to the corresponding 
limits on the CP-even couplings.  These are the most restrictive limits
available from the $ee\gamma$ and $\mu\mu\gamma$ final states.
Though the studies have equivalent sensitivities, limits from this 
analysis are less 
restrictive than those of Ref.~\cite{D0vvg} because 
of the observed event yields and $E_T^{\gamma}$ spectra.

Combining these results with our previous measurements~\cite{D095,D0vvg} 
yields 95\% CL 1D limits 
\[ |h^{Z}_{30}| < 0.36, \;   |h^{Z}_{40}| < 0.05 \;\;(h^{\gamma}_{i} = 0) \]
\[ |h^{\gamma}_{30}| < 0.37,\; |h^{\gamma}_{40}| < 0.05 \;\;
(h^{Z}_{i} = 0) \] for $\Lambda = 750$ GeV. 
These combined limits are 20\% tighter than the previous most 
restrictive combined limits~\cite{D0prd}.
Figures~\ref{fig.zzg}(c) and \ref{fig.zzg}(d) show the two-dimensional limits 
on the $ZZ\gamma$ and $Z\gamma\gamma$ couplings from the combined analyses. 

%\section{Conclusions}

In conclusion, a search for anomalous $Z$-photon couplings
was performed by studying $ee\gamma X$ and $\mu\mu\gamma  X$ 
production using the D{\O} detector.
A total of 14 (15) $ee\gamma X$ $(\mu\mu\gamma  X)$ 
candidate events were observed, in agreement 
with the $13.2 \pm 1.3$ $(16.3 \pm 2.0)$ signal events predicted by the SM
and the expected background of $1.81 \pm 0.54$ $(3.57\pm0.68)$ events.  
The photon transverse energy spectrum, the dilepton invariant mass, and the
$\ell\ell\gamma$ invariant mass are as expected from the predictions of the 
SM and provide evidence of $Z\gamma$ pair production. 
Limits on anomalous $ZZ\gamma$ and $Z\gamma\gamma$ couplings were derived. 
These results, combined with our previous measurements, provide 
the most stringent constraints on anomalous $ZZ\gamma$ and $Z\gamma\gamma$ 
couplings available.

%\section{Acknowledgments}

We thank U. Baur for providing the MC event generator and helpful discussions.
We thank the staffs at Fermilab and collaborating institutions for their
contributions to this work, and acknowledge support from the 
Department of Energy and National Science Foundation (U.S.A.),  
Commissariat  \` a L'Energie Atomique (France), 
State Committee for Science and Technology and Ministry for Atomic 
   Energy (Russia),
CNPq (Brazil),
Departments of Atomic Energy and Science and Education (India),
Colciencias (Colombia),
CONACyT (Mexico),
Ministry of Education and KOSEF (Korea),
and CONICET and UBACyT (Argentina).

%\twocolumn

\clearpage   
%\vskip -1cm

\begin{table}[h]
\begin{center}
\begin{tabular}{lcc}
Channel                          &$ee\gamma$ &$\mu\mu\gamma$ \\ \hline
PDF choice, $Q^{2}$, $k$-factor  &  6\%      & 6\%           \\
$p_{T}^{\ell\ell\gamma}$         &  1\%      & 1\%           \\
$\ell\ell\gamma$ selection efficiency  
                                 & 2.3\%     & 6.3\%         \\
Photon conversion rate           & 5\%       &  5\%          \\
Random overlap rate              & 3\%       &  3\%          \\
Luminosity                       & 5.3\%     & 5.3\%         \\ \hline
Total:                           & 10\%      & 12\%          \\
\end{tabular}
\caption{Summary of the systematic uncertainties for the predicted
$p\bar{p} \rightarrow ee \gamma $ and $\mu\mu\gamma$ signals.}
\end{center}
\label{tabl.sys}
\end{table}

\end{document}